# Experimental Investigation of Spin Transport Properties in Silicon by Using a Non-local Geometry

**Masashi Shiraishi, Yoshiya Honda, Eiji Shikoh, Yoshishige Suzuki, Teruya Shinjo**

Graduate School of Engineering Science, Osaka University. 560-8531 Toyonaka, Japan.

**Tomoyuki Sasaki, Tohru Oikawa, Kiyoshi Noguchi**

SQ Research Center, TDK Corporation, Nagano 385-8555, Japan

**Toshio Suzuki**

AIT, Akita Prefectural R&D Center (ARDC), Akita 010-1623, Japan

E-mail) shiraishi@ee.es.osaka-u.ac.jp

**Abstract**

A systematic investigation of spin transport properties in silicon at 8 K by using a non-local geometry is presented. The spin injection signal in the non-local scheme is found to increase in proportion to the evolution of bias electric currents. Theoretical fittings using the Hanle-type spin precession signals reveal that the spin polarization of the transported spin in the Si is much less affected by the change in the bias electric current compared with a case of the other spin devices, which induces a unique bias dependence of the spin signals.

Spintronics using Si has been an emerging research field in recent years, because Si is a ubiquitous and toxic-free material possessing a lattice inversion symmetry that yielding good spin coherence. In addition, the existing infrastructures for Si-based electronics will be used for fabricating Si-based spintronic devices. Spin injection, transport and manipulation in Si allow us to envisage novel beyond-CMOS devices, such as Sugahara-Tanaka type spin transistors [1]. Much effort has been devoted, and various important milestones have already been reached, such as hot electron spin injection [2], extremely long spin coherence under application of a bias voltage [3], spin transport by using a non-local technique at low temperature [4] and at room temperature (RT) [5], and electrical spin accumulation at RT [6]. However, previous studies have mainly shed light on spin injection and the measurement of spin coherence, and many aspects of the physics of spin transport in Si remain unknown.

In this letter, we report on a detailed study of spin transport properties in Si focusing on spin signals and spin polarization under biasing conditions. A systematic investigation of non-local magnetoresistance (MR) measurements shows that the spin signal intensity is almost proportional to the bias electric current, which has not been achieved in conventional metallic and inorganic semiconductor spin devices. In addition, Hanle-type spin precession measurements reveal that the spin polarization is robust for the bias current.

The present device consists of a P-doped ($\sim 5 \times 10^{19}$ cm$^{-3}$) n-type Si channel, with a width and thickness of 21 μm and 80 nm, respectively, equipped with two ferromagnetic (FM) and two non-magnetic (NM) electrodes (see Fig. 1(a)), all fabricated onto a SOI (Silicon-on-Insulator) substrate. The FM electrodes were Fe (13 nm) with a MgO tunneling barrier (0.8 nm), and their dimensions were $0.5 \times 21$ μm$^2$ (FM1) and $2 \times 21$ μm$^2$ (FM2), respectively. The center-to-center length between the FM electrodes was 1.86 μm. The NM electrodes were made of Al. All electrodes were covered by Ta (10 nm)/Cu (20 nm)/Ta (10 nm) coating to prevent oxidation. Details of the fabrications process is described elsewhere [5]. The resistance of the Si channel between FM1 and NM1 (the current injection circuit) at 8 K was determined by using a 4-terminal measurement, where the resistance was decreased from 930 to 550 Ω as the bias electric dc current was increased. Non-local MR and Hanle-type spin precession measurements [7] were performed at 8 K by applying an external magnetic field to the *y*-axis and *z*-axis, respectively. Measurements were typically repeated four times by using a four-probe system (ST-500, Janis Research) and Physical Property Measurement System (PPMS, Quantum Design), and the data obtained were averaged to enable a reliable theoretical fitting. The non-local MR measurement was carried out under a dc electric current injection, where the spins were injected from the FM into the Si, in order to observe the bias dependence of the spin signals precisely. The applied dc electric current was varied from 0.5 to 4 mA, which corresponds to a bias voltage of 0.42 V to 2.2 V. A stronger electric current was not applied to prevent breakdown of the sample.

Figure 2(a) shows a typical non-local MR signal observed in the Si spin device at 8 K, where an apparent hysteresis of the non-local resistance was observed. This implies that spins were definitely injected and transported in the Si channel. The non-local spin voltage, $\Delta V_{non-local}$, was defined as the difference of the output voltages under parallel and anti-parallel magnetization alignments of FM1 and FM2, as shown in Fig. 2(a). The MgO tunneling barrier was inserted between the Fe and Si, thus the intensity of the spin signal is expressible as [7],

$$\Delta V_{non-local} = \frac{P^2 \lambda_{sf}}{2\sigma A} \cdot I_{inject} \exp(-\frac{L}{\lambda_{sf}}), \qquad (1)$$

where $P$ is spin polarization, $\lambda_{sf}$ is spin diffusion length, $I_{inject}$ is electric current, $L$ is the gap length between the two FM electrodes, $\sigma$ and A are the conductivity and thickness of the Si channel, respectively. The bias dependence of the spin signals is shown in Fig. 2(b), where an unprecedented behavior was observed. The spin signals increased almost linearly as the dc bias current was increased up to 1.5 mA, and deviated from the linear dependence above 2 mA. However, the deviation was not so obvious; the spin signal at 4 mA exhibited a 21% decrease from the expected linear dependence. Bias dependence of spin signals has been vigorously investigated in tunnel magnetoresistance (TMR) devices [8], giant magnetoresistance (GMR) devices with tunneling barriers [9] and Fe/GaAs spin valves [10], where the spin signals became a half of the maximum (50% decrease from a expected linear dependence) under applications of bias voltages of 1 V at RT [8], 0.4 V at RT [9] and 0.2 V at 10 K [10], respectively. Experimentally, the spin signals decreased rapidly and monotonously as the bias current evolved, which is thought to be attributable to spin scattering and loss of spin polarization of transported spins at an interface between the FM and the spin channel (or the tunneling barrier). The finding of our study reveals a dramatic improvement in terms of the bias dependence of the spin signals because the spin signal was still 79% of the maximum value even when the bias voltage was 2.2 V was applied, and Eq. (1) suggests that the linear dependence is ascribable to the robustness of the spin polarization of the spins transported into the Si [11]. A theoretical study revealed that a linear dependence of the spin signals on the bias current can be achieved [12] where the spin signal exhibits a linear dependence in the case of spin injection into degenerated semiconductors if spin drift motion can be ignored. The present case fits this description exactly and our experimental observation qualitatively corresponds to the theoretical prediction. However, the bias current dependence of the spin signal does not sufficiently ensure the robustness. Hanle-type spin precession experiments are among the most promising tools for precise estimations of spin polarization and spin coherence. Hence, systematic measurements of Hanle-type spin precession were conducted as the dc bias current was varied.

Figures 3(a)-(h) show the spin signals obtained with the Hanle-type spin precession under various bias current conditions. The motion of a spin can be described by the drift-diffusion equation,

$$\frac{\partial S(x,t)}{\partial t} = D\frac{\partial S^2(x,t)}{\partial x^2} - v\frac{\partial S(x,t)}{\partial x} - \frac{S(x,t)}{\tau_{sf}},$$ where $S(x,t)$ is spin density at a position ($x$) and time ($t$), $D$ is the diffusion constant, $\tau_{sf}$ is the spin relaxation time and $v$ is the spin drift velocity which is negligible in the present experimental scheme. Using the above equation, an intensity of a spin signal with the Hanle-type spin precession can be derived by using a Green's function of the equation. Thus, the analytical equation describing the non-local spin resistance with Hanle-type spin precession becomes [13]

$$\frac{V_{non-local}}{I_{inject}} = \frac{P^2 \lambda_N}{2\sigma A}\exp(-\frac{L}{\lambda_N})(1+\omega^2\tau_{sf}^2)^{-1/4}\exp(-\frac{L}{\lambda_N})\{\sqrt{\frac{1}{2}(\sqrt{1+\omega^2\tau_{sf}^2}+1}-1\} \\ \times \cos\{\frac{\tan^{-1}(\omega\tau_{sf})}{2}+\frac{L}{\lambda_N}\sqrt{\frac{1}{2}(\sqrt{1+\omega^2\tau_{sf}^2}-1)}\}, \quad (2)$$

where $\omega$ is the Larmor frequency. It should be noted that the above equation is valid in the limit where the width of FM electrodes is negligible in comparison with the spin diffusion length. Thus, Eq. (2) was further corrected by taking into account the width, which will be described elsewhere [14], and the corrected equation was used for the theoretical fittings under an assumption that the electric current was uniformly injected from the electrode to the Si. The results of the estimated spin polarization and spin coherence are shown in Fig. 4. The spin relaxation time, spin diffusion length and the spin polarization at 8 K were estimated to be roughly 10 ns, 2 μm and 5%, respectively. The spin diffusion length and the spin relaxation time showed no obvious decrease with increasing bias current. The spin polarization was characterized by little variation below 1.5 mA and a gradual decrease above 2 mA, similarly to the dependence of the spin signal on the bias current. This finding implies that the slight deviation from the linearity in the spin signal (see Fig. 2(b)) is attributable to the loss of the spin polarization, and the robustness of the spin polarization was experimentally verified. The spin polarization estimated at 4 mA decreased ca. 26% compared with the value estimated at 0.5 mA, which has comparably good accordance with the value of the decrease of the spin signal (ca. 21%) as estimated above. Hence, it is elucidated that the decrease of the spin signal is attributable to the loss of the spin polarization, in other words, the much better robustness of the spin signal on the bias current is ascribable to the robustness of the spin polarization of the spins transported in Si. A similar linear dependence of the spin signals on bias electric current has been reported in graphene spin valves [15, 16]. One may conclude this robustness to be a spin transport feature unique to group-IV elements, although a systematic study has not yet been conducted. However, the appearance of robustness was only speculation in the case of graphene. Our experimental study is the first to reveal the detailed spin transport properties of Si and to yield experimental evidence of the robustness of the spin polarization in Si, which is a good material for further discussion of the physics of spin transport in group-IV elements. From the stand point of practical applications, the discovery of this robustness can be an important milestone in operating spin transistors because the spin signal is not diminished under the application of a bias voltage.

In summary, we conducted a systematic study using non-local spin transport and Hanle-type spin precession measurements, and found that the spin signals exhibited a linear dependence on the bias current. This experimental finding was attributed to the robustness of the spin polarization of the spins injected and transported into Si, which was experimentally verified by the Hanle-type spin precession analyses.

The authors thank T. Sugimura and K. Muramoto for their assistance with experimental set-up and measurements.

**Figure captions**

**Figure 1** (Color online, Not to scale)

Schematic of the Si spin device. The highly doped ($5\times10^{19}$ cm$^{-3}$) Si channel is equipped with four electrodes (two NM, two FM). The width of the FM electrodes was varied to control the coercive force. An electric current is injected into one circuit (NM1-Si-FM1), and an output spin accumulation voltage is detected by the other circuit (FM2-Si-NM2). The two circuits are completely separated, which allows us to eliminate spurious signals, such as the anisotropic MR effect. The external magnetic field was applied to the *y*-axis in the non-local MR measurements and to the *z*-axis in Hanle-type spin precession measurements. All measurements were carried out at 8 K.

**Figure 2** (Color online)

(a) Typical spin injection signal in the non-local scheme at 8 K, where the bias electric current was set to 2 mA. The blue and red solid lines show the spin signals for forward and backward sweeping of the external magnetic field, respectively.

(b) Bias current dependence of the non-local spin signals. The spin signal increased in proportion to the bias current up to 1.5 mA, and the slight deviation from the linear dependence can be seen above 2 mA. The dashed line is a fitting line for experimental data below 1.5 mA, which was obtained by using a root mean square method.

**Figure 3** (Color online)

Hanle-type spin precession signals observed in the Si spin device at 8 K. The vertical axis is the non-local resistance, defined as (spin voltage)/ (injection bias current). The bias current was set to (a) 0.5 mA, (b) 1 mA, (c) 1.5 mA, (d) 2 mA, (e) 2.5 mA, (f) 3 mA, (g) 3.5 mA and (h) 4 mA. The red closed squares are the measurement results and blue solid lines are theoretical fitting lines.

**Figure 4** (Color online)

Injection bias current dependence of spin relaxation time (upper panel), spin diffusion length (middle panel) and spin polarization (lower panel) at 8 K. These spin transport parameters were estimated by using the results of the Hanle-type spin precession. The red dashed line shows the bias current regime where the spin signals did not show a linear dependence.

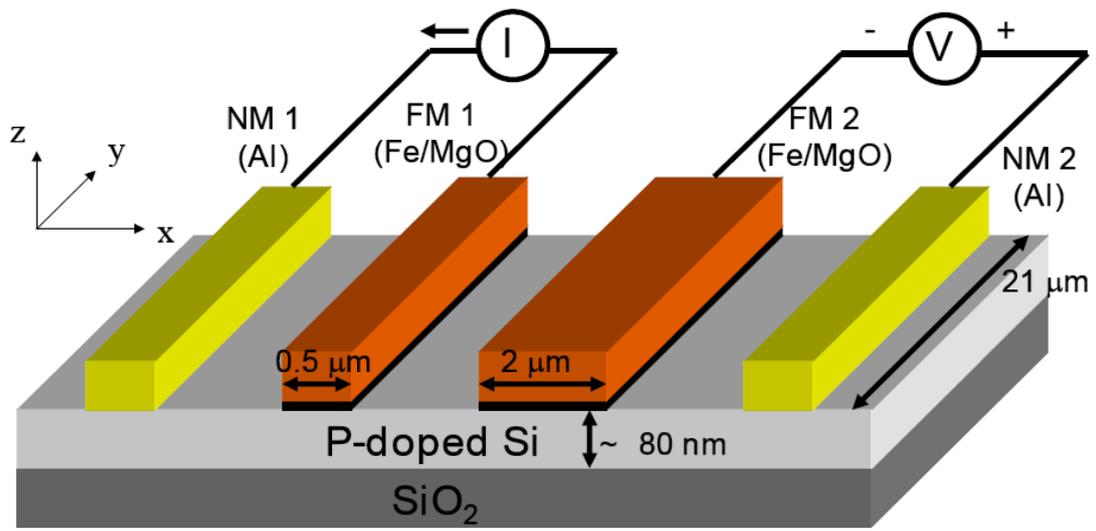

Fig. 1 Shiraishi et al.

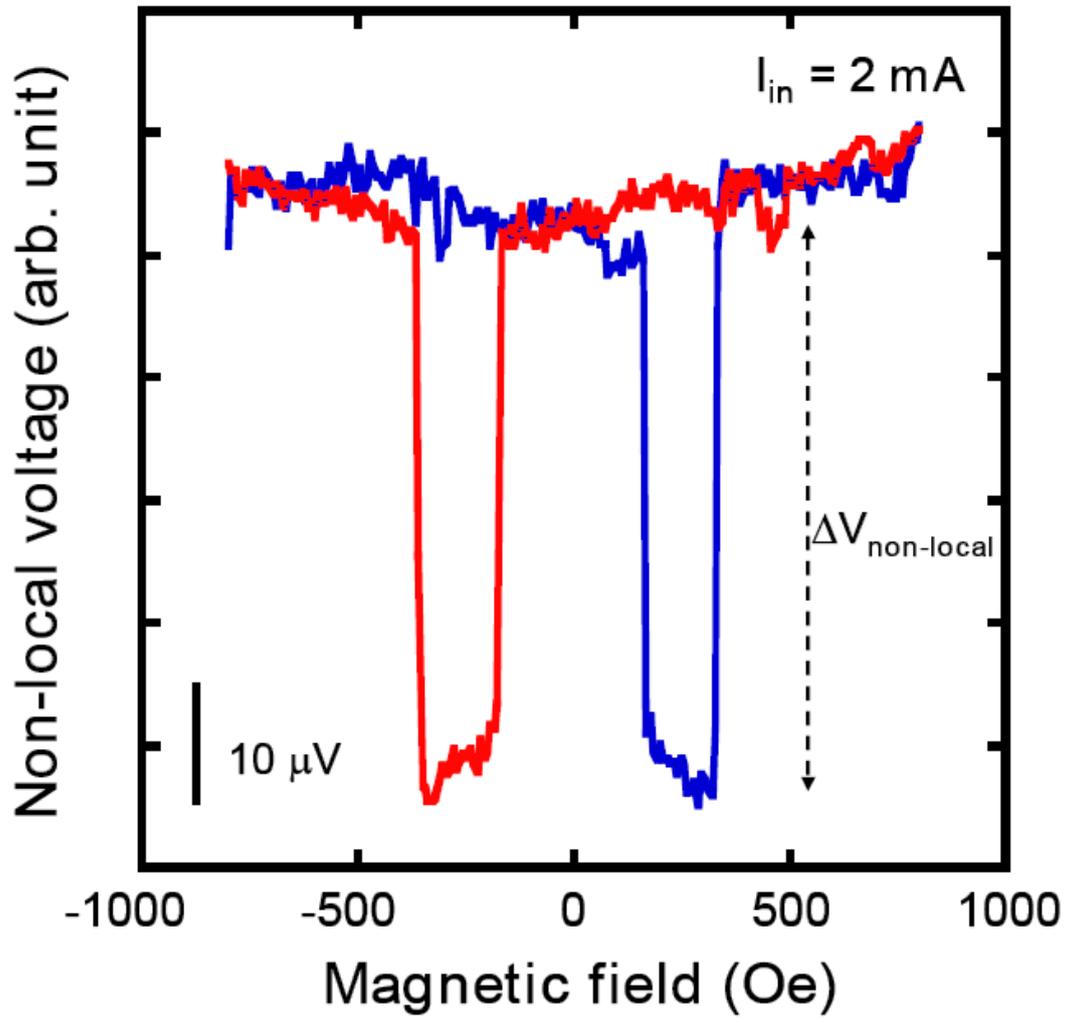

Fig. 2(a) Shiraishi et al.

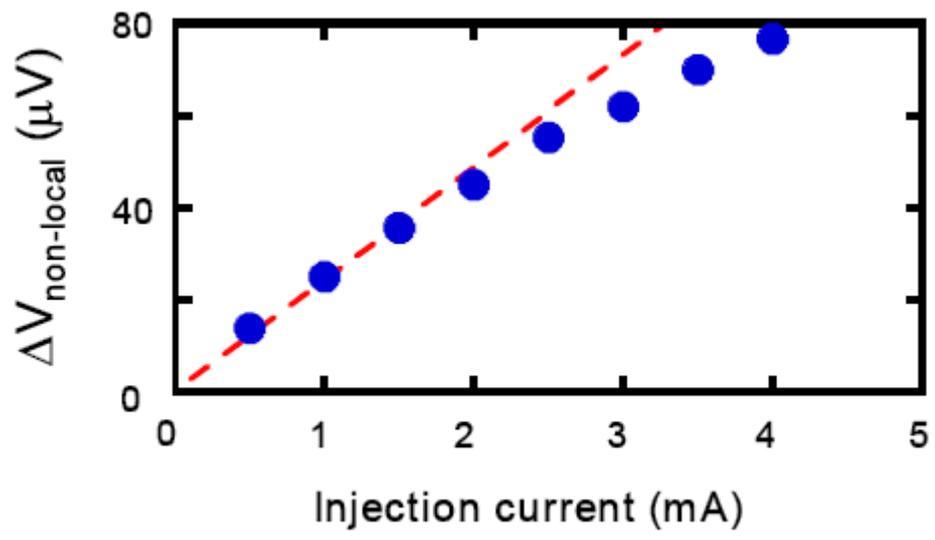

Fig. 2(b) Shiraishi et al.

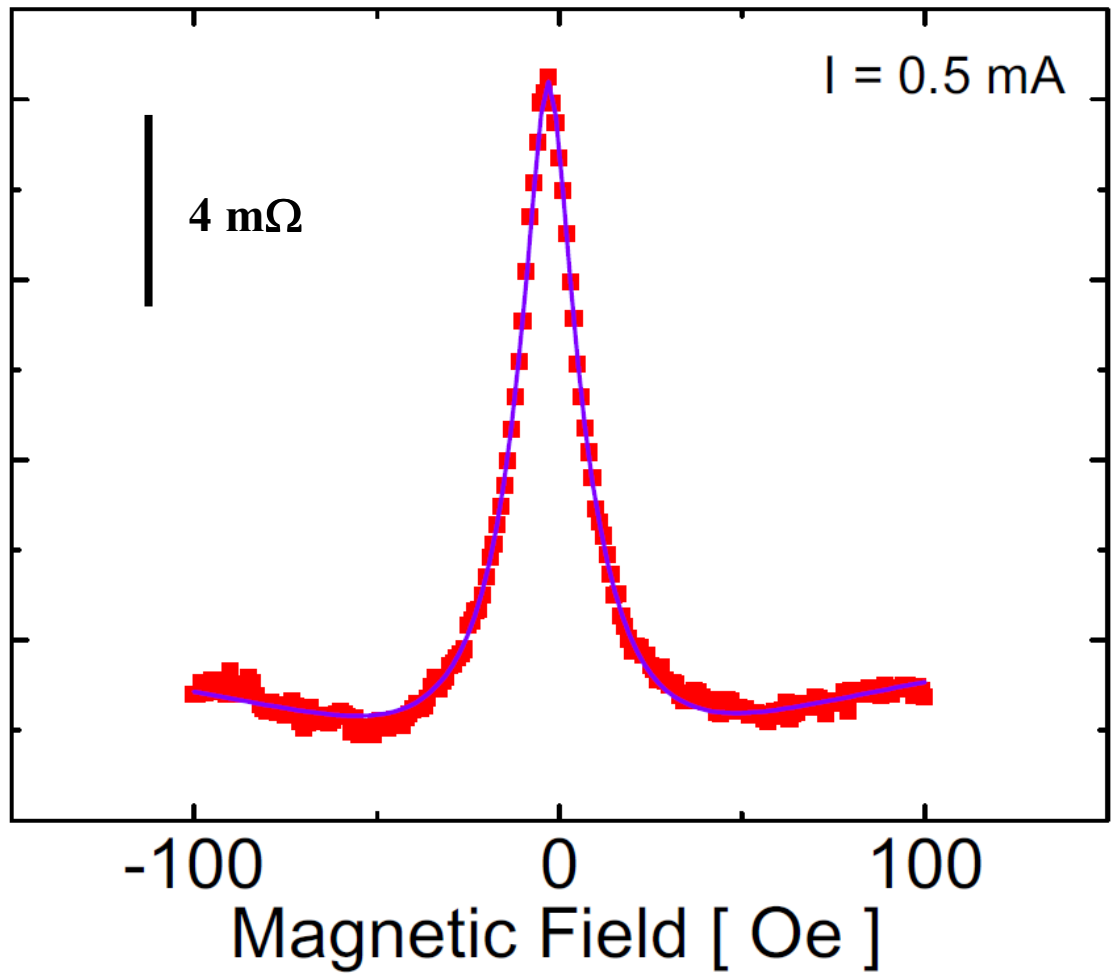

Fig. 3(a) Shiraishi et al.

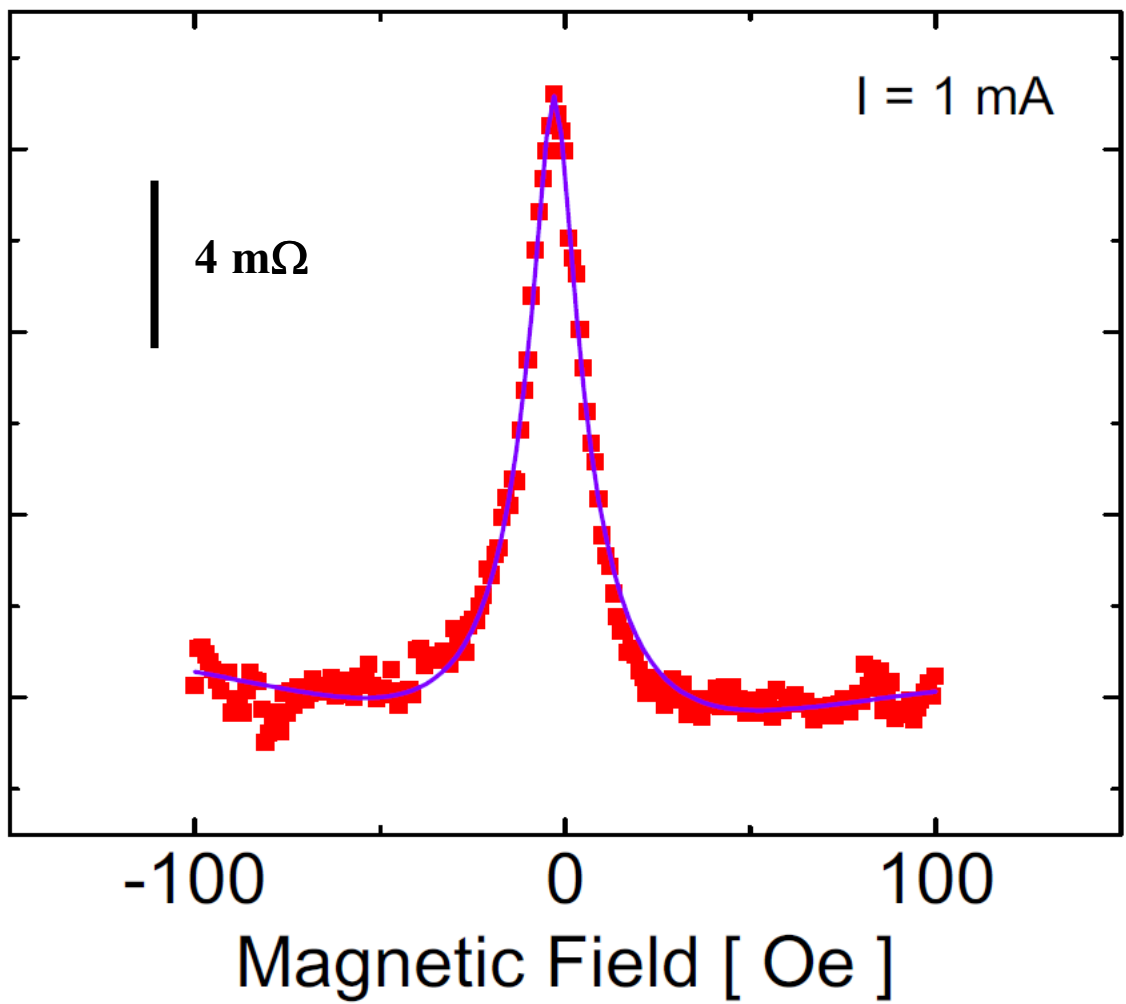

Fig. 3(b) Shiraishi et al.

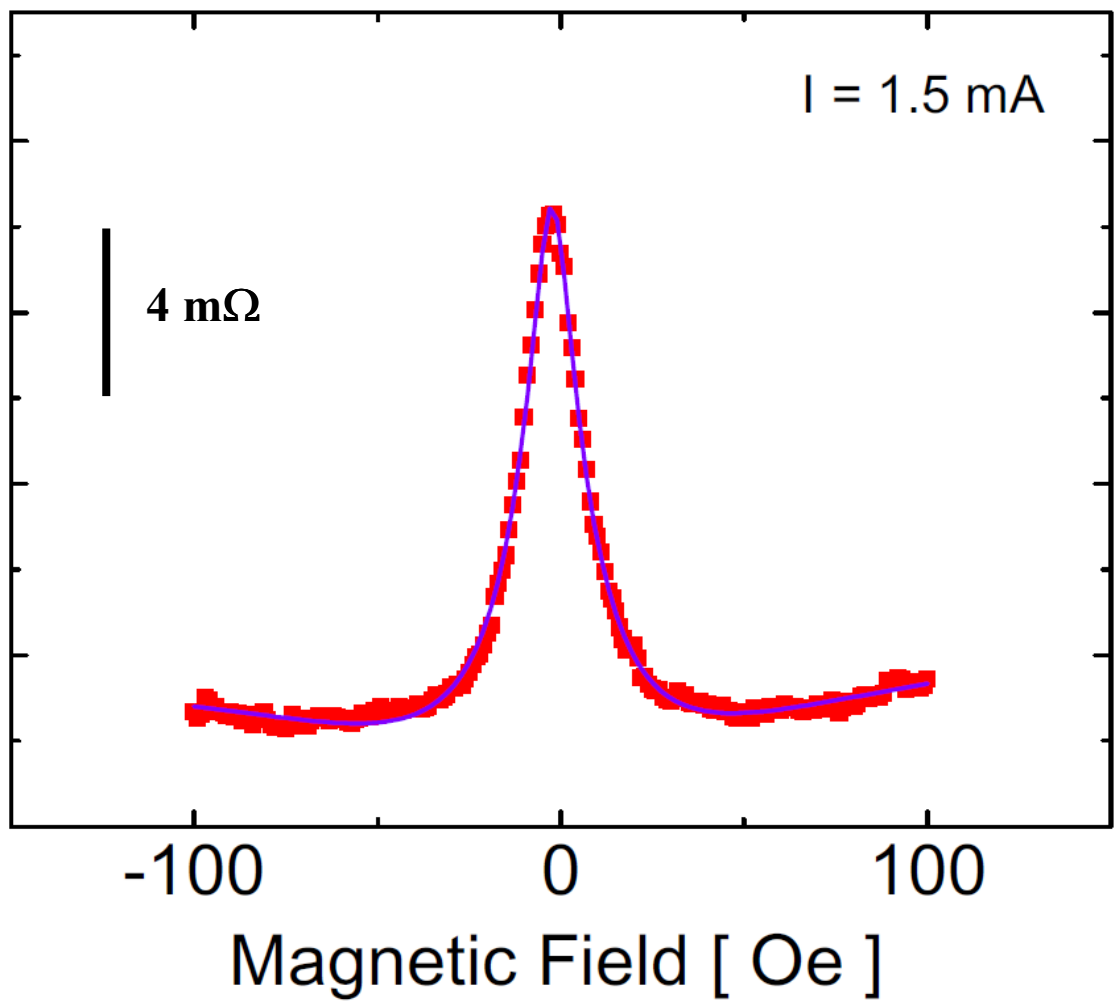

Fig. 3(c) Shiraishi et al.

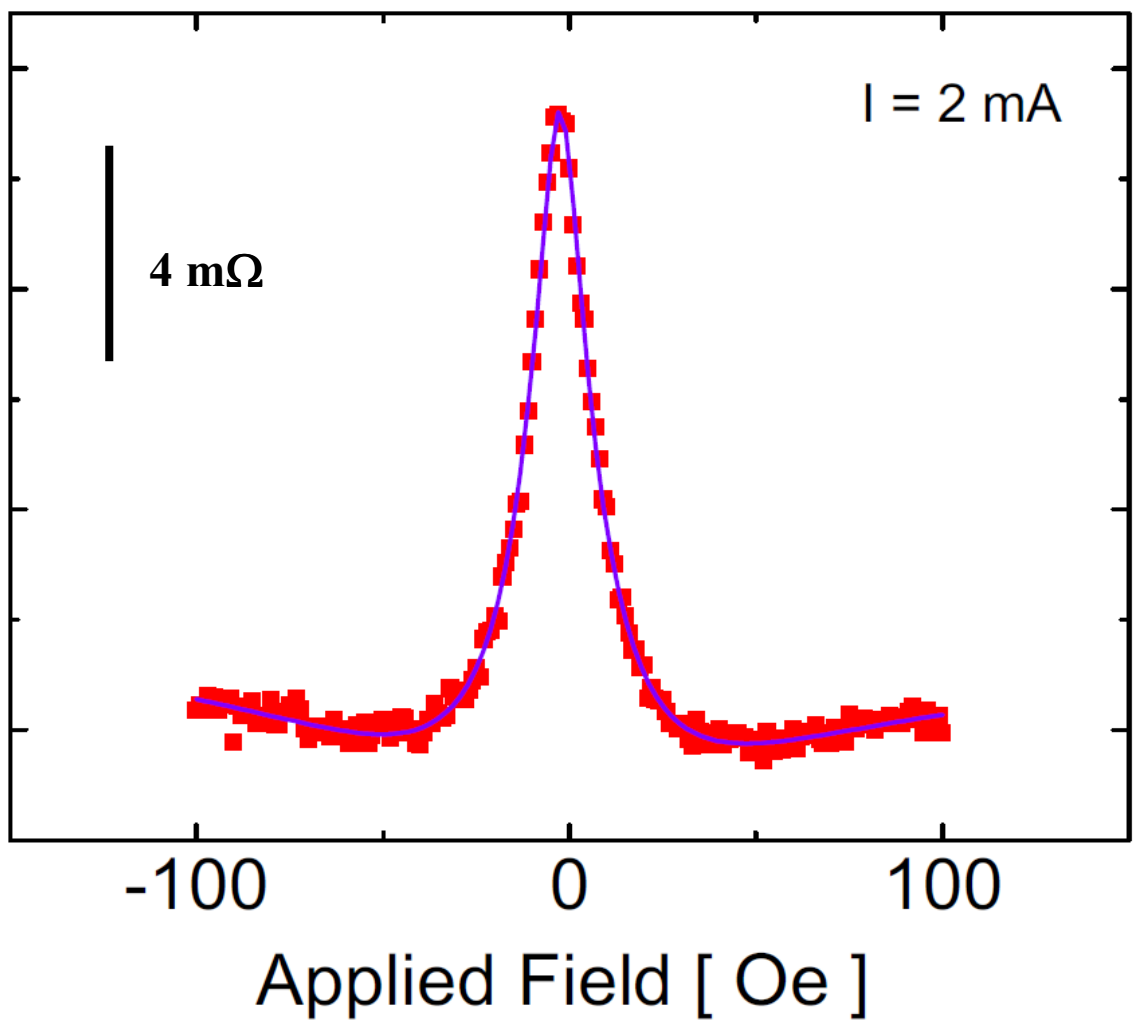

Fig. 3(d) Shiraishi et al.

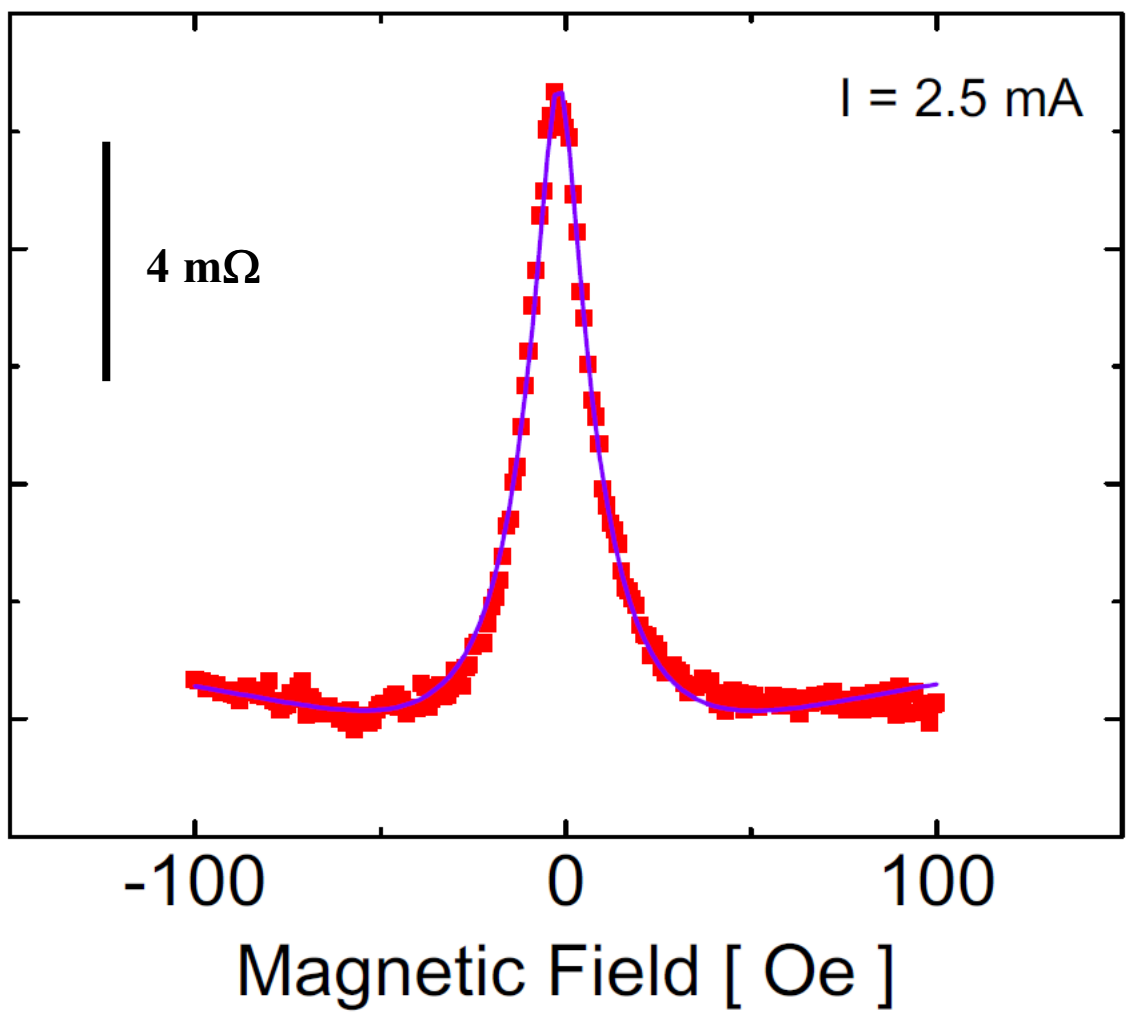

Fig. 3(e) Shiraishi et al.

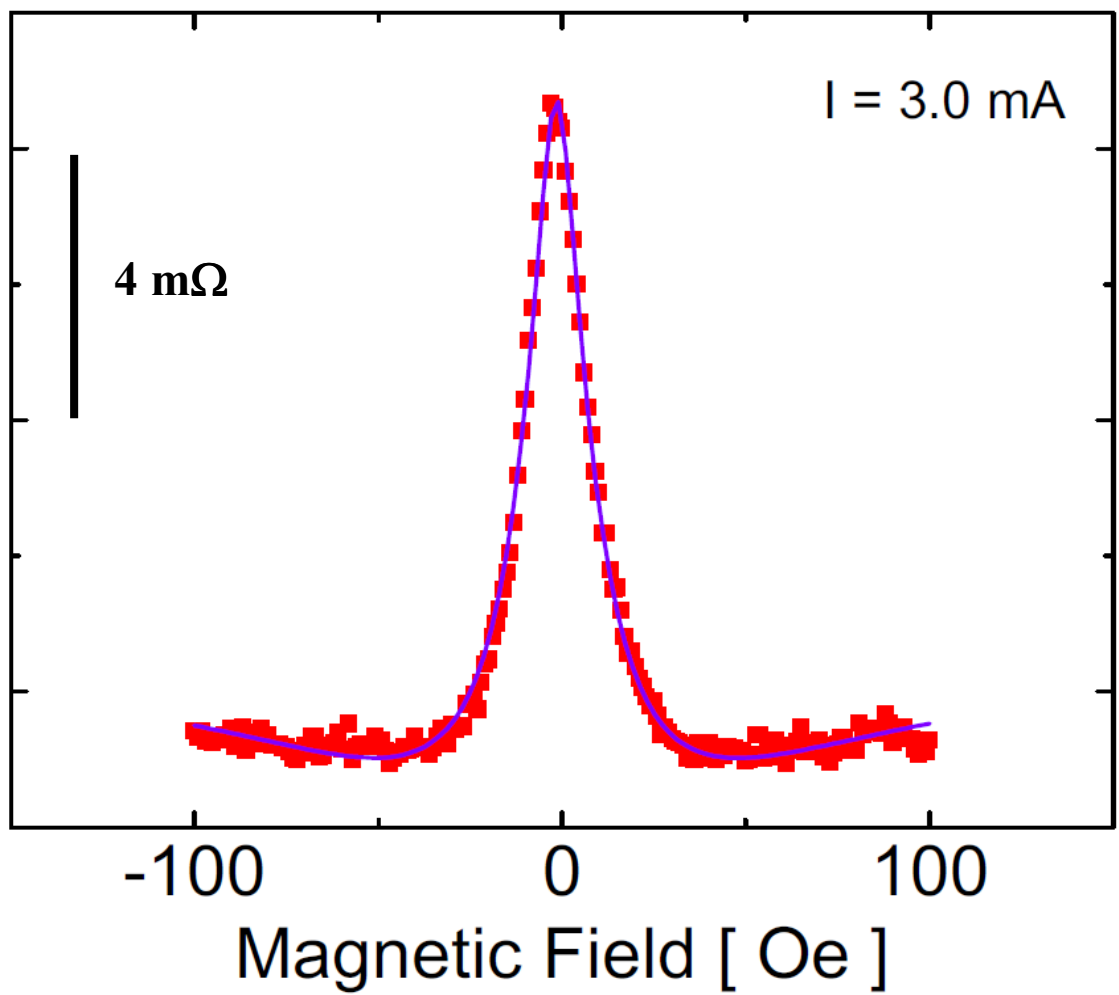

Fig. 3(f) Shiraishi et al.

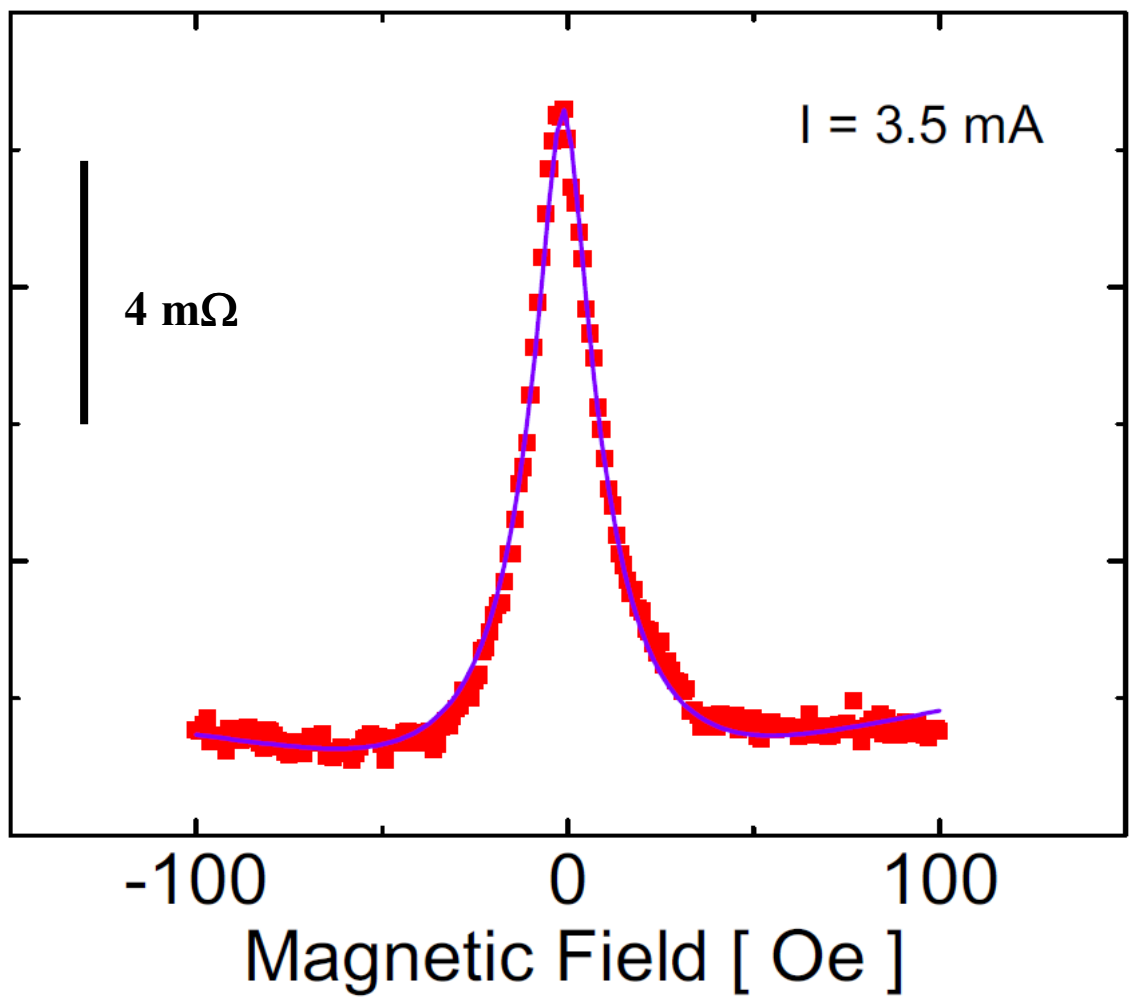

Fig. 3(g) Shiraishi et al.

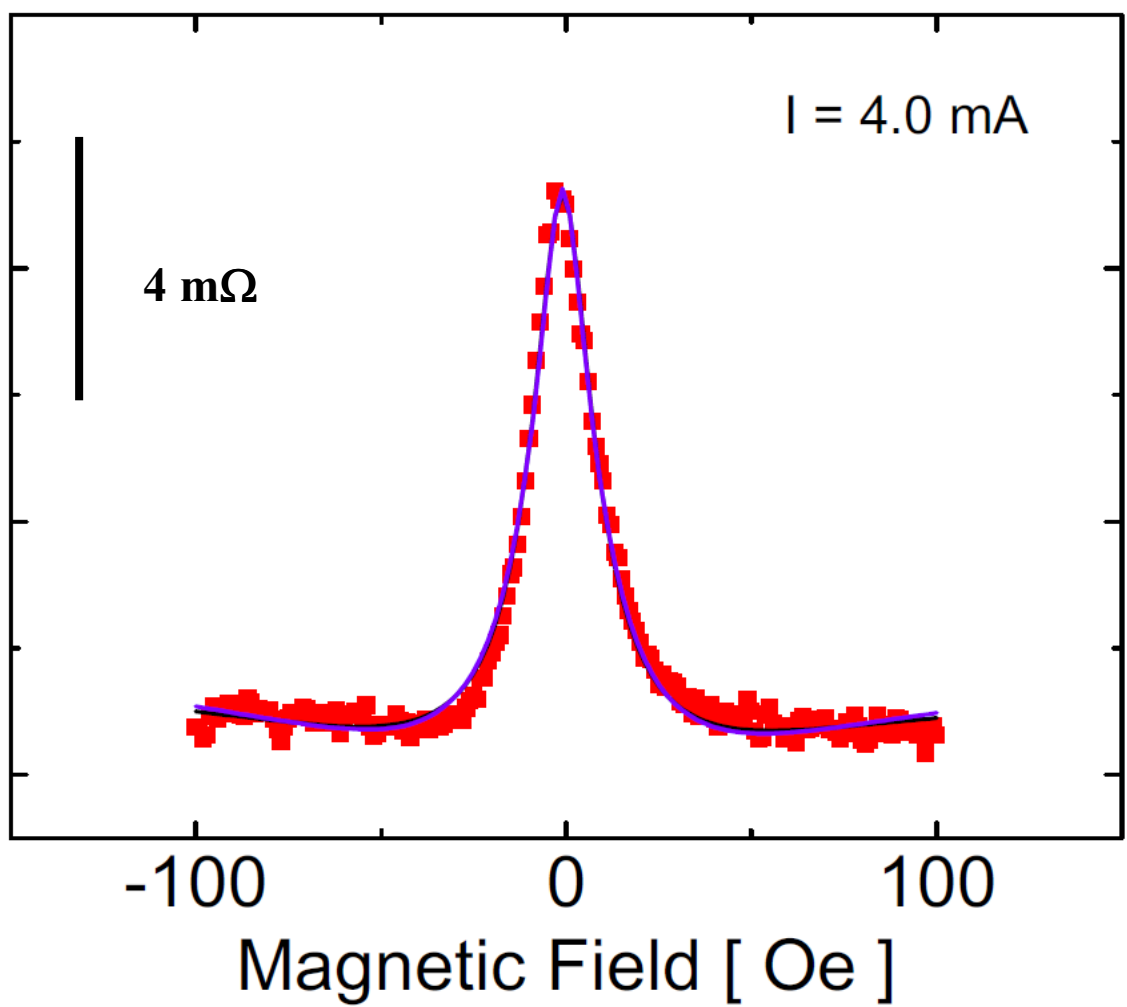

Fig. 3(h) Shiraishi et al.

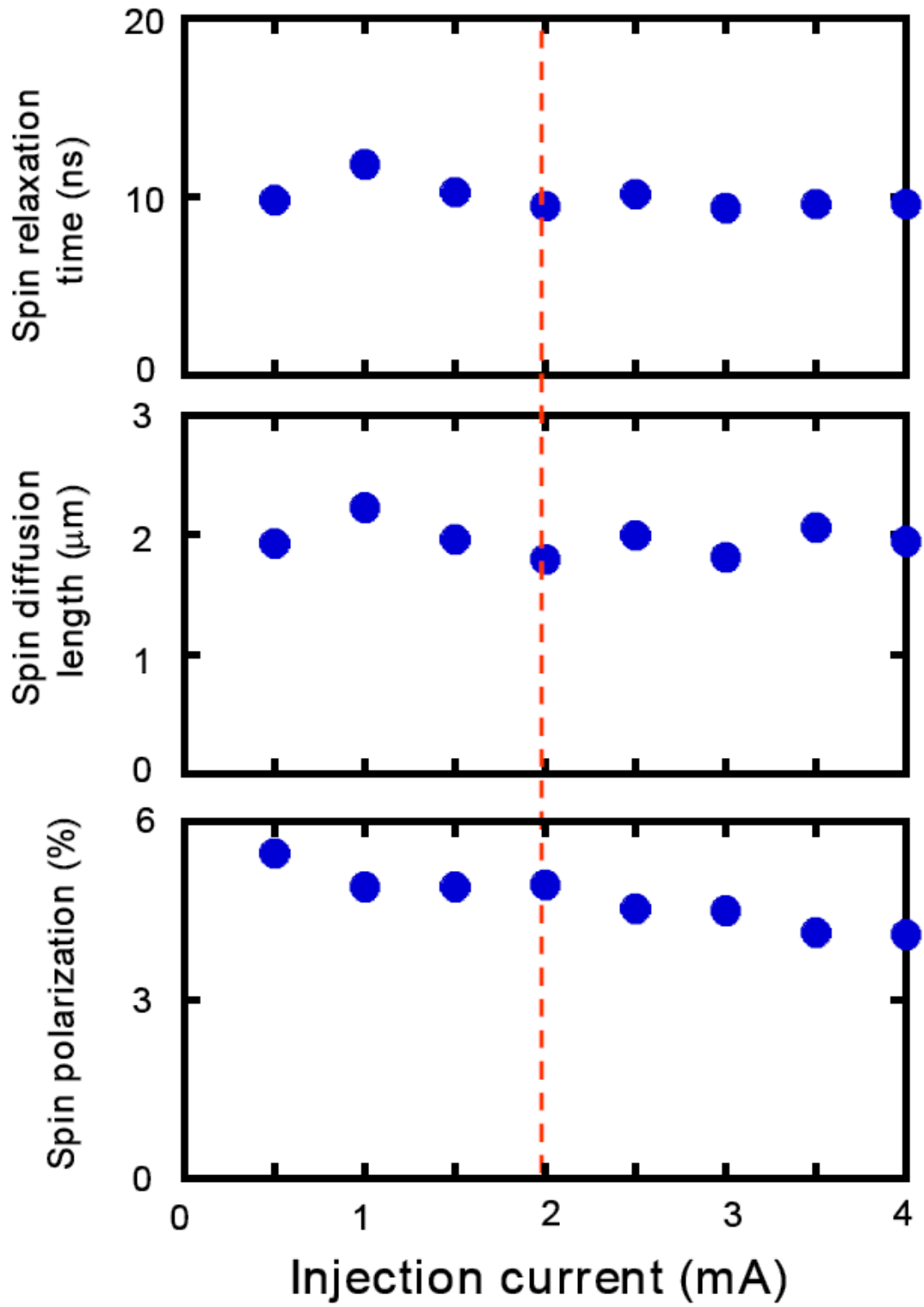

Fig. 4 Shiraishi et al.